\documentclass[twocolumn,prb,superscriptaddress,showpacs]{revtex4-1}
\usepackage{amsmath,amssymb}
\usepackage{graphicx}
\usepackage{verbatim}
\usepackage{hyperref}
\usepackage{amsfonts}
\usepackage{dcolumn}
\usepackage{bm}

\begin{document}

\title{\large \bf Symmetry protected topological orders of 1D spin systems
with $D_2+T$ symmetry }

\author{Zheng-Xin Liu}
\affiliation{Institute for Advanced Study, Tsinghua University,
Beijing, 100084, P. R. China}
\affiliation{Department of Physics,
Massachusetts Institute of Technology, Cambridge, Massachusetts
02139, USA}

\author{Xie Chen}
\affiliation{Department of Physics, Massachusetts Institute of
Technology, Cambridge, Massachusetts 02139, USA}

\author{Xiao-Gang Wen}
\affiliation{Department of Physics, Massachusetts Institute of
Technology, Cambridge, Massachusetts 02139, USA}
\affiliation{Institute for Advanced Study, Tsinghua University,
Beijing, 100084, P. R. China}

\begin{abstract}
In [Z.-X. Liu, M. Liu, X.-G. Wen, arXiv:1101.5680], we studied 8
gapped symmetric quantum phases in $S=1$ spin chains 
which respect a discrete spin rotation $D_2 \subset SO(3)$ and time
reversal $T$ symmetries. In this paper, using a generalized
approach, we study all the 16 possible gapped symmetric quantum
phases of 1D integer spin systems with only $D_2+T$ symmetry. Those
phases are beyond Landau symmetry breaking theory and cannot be
characterized by local order parameters, since they do not break any
symmetry. They correspond to 16 symmetry protected topological (SPT)
orders.  We show that all the 16 SPT orders can be fully
characterized by the physical properties of the symmetry protected
degenerate boundary states (end `spins') at the ends of a chain
segment. So we can measure and distinguish all the 16 SPT orders
experimentally. We also show that all these SPT orders can be
realized in $S=1$ spin ladder models. The gapped symmetric phases
protected by subgroups of $D_2+T$ are also studied.  Again, all
these phases can be distinguished by physically measuring their end
`spins'.

\end{abstract}
\pacs{75.10.Pq, 64.70.Tg }

\maketitle

\section{introduction}

In recent years, topological order\cite{Wtop,Wrig} and symmetry protected
topological (SPT) order\cite{Wqoslpub,GWtefr} for gapped quantum ground states
has attracted much interest. Here `topological' means that this new kind of
orders is different from the symmetry breaking
orders.\cite{L3726,GL5064,LanL58}  The new orders include fractional quantum
Hall states\cite{TSG8259,L8395}, 1D Haldane phase\cite{H8364}, chiral spin
liquids,\cite{KL8795,WWZ8913} $Z_2$ spin liquids,\cite{RS9173,W9164,MS0181}
non-Abelian fractional quantum Hall states,\cite{MR9162,W9102,WES8776,RMM0899}
quantum orders characterized by projective symmetry group
(PSG),\cite{Wqoslpub,W0303a} topological
insulators\cite{KM0501,KM0502,BZ0602,MB0706,FKM0703,QHZ0824}, etc.

Recent studies indicate that the 
patterns of 
entanglements provide 
a systematic and comprehensive  point of view 
to understand topological orders and SPT orders.\cite{LW0605,KP0604,
CGW1038, W0275} The phases with long-ranged entanglement have intrinsic
topological orders, while symmetric short-range entangled
nontrivial phases are said to have SPT orders. With a definition of phase and
phase transition using local unitary transformations, one can get a
complete classification for all 1D gapped quantum
phases,\cite{CGW1107,SPC1032,CGW1123} and partial classifications
for some gapped quantum phases in higher
dimensions.\cite{LWstrnet,CGW1038,GWW1017}


In contradiction to the suggestion from the symmetry breaking
theory, even when the ground states of two Hamiltonians have the
same symmetry, sometimes, they still cannot be smoothly connected by
deforming the Hamiltonian without closing the energy gap and causing
a phase transition, as long as the deformed Hamiltonians all respect
the symmetry.  So those two states with the same symmetry can belong
to two different phases. Those kind of phases, if gapped, are called
SPT phases. The Haldane phase of spin-1 chain\cite{H8364} is the
first example of SPT phase, which is known to be protected by the
$D_2=\{E, R_x=e^{i\pi S_x}, R_y=e^{i\pi S_y}, R_z=e^{i\pi S_z}\}$
symmetry.\cite{PBT0959} Interestingly, when additional time reversal
symmetry is present, more SPT phases emerges.\cite{LLW1180,CGW1123}

Topological insulators\cite{KM0501,BZ0602,KM0502,MB0706,FKM0703,
QHZ0824} is another examples of SPT phases which has attracted much
interest in literature. Compared to the topological insulators
formed by free electrons, most SPT phases (including the ones
discussed in this paper) are strongly correlated.  A particular kind
of strongly correlated SPT phases  protected by time reversal
symmetry is called the fractionalized topological insulators by some
people.

An interesting and important question is how to classify different
1D SPT phases even in presence of strong correlations/interactions.
For the Haldane phase in spin chains, it was thought that the
degenerate end states and non-local string order can be used to
describe the hidden topological order. However, if we remove the
spin rotation symmetry but keep the parity symmetry, the Haldane
phase is still different from the $\otimes_i|z\rangle_i$
($S_z|z\rangle =0$) trivial phase, despite that the degenerate end
states and non-local string order are destroyed by the absence of
spin rotation symmetry.\cite{BTG0819,GWtefr,PBT0959}

Recently, it was argued in \Ref{LH0804} that the entanglement
spectrum degeneracy (ESD) can be considered as the criteria to tell
whether a phase is topologically ordered or not.  However, it is
known that all 1D gapped states are short range entangled and have
no intrinsic topological orders from entanglement point of
view.\cite{VCL0501,CGW1107} On the other hand, many gapped 1D phases
have non-trivial ESD.  So ESD cannot correspond to the intrinsic
topological orders.  Then, one may try to use ESD to characterize
non-trivial SPT orders as suggested in \Ref{PBT1039}.
ESD does appear to describe non-translation invariant SPT phases protected by on-site symmetry.
In particular, the ESD reveal an important connection to the
projective representation of the on-site symmetry group.\cite{PBT1039}

It turns out that a clear picture and a systematic classification of all 1D SPT
phases can be obtained after realizing the deep connection between local
unitary transformation and gapped (symmetry protected) topological
phases.\cite{CGW1107,SPC1032,CGW1123} In particular, for 1D systems, all gapped
phase that do not break the symmetry are classified by the 1D representations
and projective representations of the symmetry group $G$ (\ie by the group
comology classes $\cH^1[G,U_T(1)]$ and $\cH^2[G,U_T(1)]$, see appendix
\ref{app:cohomology}).\cite{CGW1107,SPC1032,CGW1123}

In our previous paper, we have calculated the eight classes of
unitary projective representations of the point group
$D_{2h}=D_2+T$, based on which we predicted eight SPT phases in
integer spin models that respect the $D_{2h}$ symmetry. We realized
four interesting SPT phases in $S=1$ spin chains, and showed that
these phases can be distinguished experimentally by their different
responses of the end states to magnetic field. In this paper we will
show that the group $D_2+T$ has totally 16 projective
representations when the representation of $T$ is anti-unitary. We
then study the properties of the corresponding 16 SPT phases, such
as the dimension of their degenerate end states and their response
to perturbations. Interestingly, we find that all these SPT phases
can be distinguished by their different responses of the end states
to various physical perturbations. We also show that all these SPT
phases can be realized in spin ladders. Finally we discuss the
situations when the symmetry reduces to the subgroups of $D_{2}+T$.

This paper is organized as following. In section \ref{sec. 16 SPT
phase} we show that there are 16 SPT phases that respect $D_2+T$
symmetry, and all these phases can be distinguished experimentally.
The realization of the 16 SPT phases in $S=1$ spin chains and spin
ladders are given in section \ref{sec. realizations}. In section
\ref{sec. sugroups}, we discuss the projective representations and
SPT phases of two subgroups of $D_2+T$. Section \ref{sec. summary}
is the conclusion and discussion. Some details about the
derivations, together with a brief introduction to projective
representations (and group cohomology) and general classification of
SPT phases, are given in the appendices.

\section{Distinguishing 16 SPT phases with $D_{2}+T$
symmetry}\label{sec. 16 SPT phase}

\begin{table*}[htbp]
\caption{All the projective representations of group $D_{2h}=D_2 +
T$. We only give the representation matrices for the three
generators $R_z,R_x$ and $T$. $K$ stands for the anti-linear
operator. The 16 projective representations corresponds to 16
different SPT phase. This result agrees with the classification of
combined symmetry $D_2+T$ given in Ref.~\onlinecite{CGW1123}. The
indexes $(\om,\bt,\ga)=(\omega(D_2), \beta(T), \gamma(D_2))$ show
this correspondence. Five of these SPT phases can be realized in
$S=1$ spin chain models and others can be realized in $S=1$ spin
ladders or large-spin spin chains. The active operators are those
physical perturbations which (partially) split the irreducible end
states. } \label{tab:prjRep2}
\begin{ruledtabular}
\begin{tabular}{c|ccc |c|c|c|c}
&$R_z$&$R_x$&$T$& $\omega,\beta,\gamma $ & dim. &active operators\footnote{In the ground states of SPT phases corresponding to the 2-dimensional projective representations, the active operators behave as $(\sigma_x,\sigma_y,\sigma_z)$, and for the 4-dimensional projective representations, the active operators behave as ($\si_x\otimes I,\si_y\otimes I,\si_z\otimes I, \si_x\otimes\si_x,$\\$\si_y\otimes\si_x,\si_z\otimes\si_x,I\otimes\si_x, \si_x\otimes\si_y,\si_y\otimes\si_y,\si_z\otimes\si_y, I\otimes\si_y,\si_x\otimes\si_z,\si_y\otimes\si_z,\si_z\otimes\si_z,I\otimes\si_z$).} & spin models ($S=1$)\\
\hline
$E_0$       &   $ 1$   &    $ 1$  &     $  K$        &\ 1,\ 1,$A$   &1& &chain(trivial phase)\\
$E_0'$      &   $  I$  &    $  I$ &     $ \sigma_y K$&\ 1,-1,$A$   &2&$(S_{xyz},S_{xyz},S_{xyz}$)\footnote{We notate $S_{mn}=S_mS_n+S_nS_m$, where $m,n=x,y,z$. For $S=1$, $S_{xyz}$ means a multi-spin operator, such as $S_{xy,i}S_{z,i+1}$ . } &ladder\\ 
\hline
$E_1$       &    $I$     & $i\sigma_z$& $ \sigma_y K$&\ 1,-1,$B_1$ &2& $(S_z,S_z,S_{xyz})$ &ladder\\
$E_1'$      &    $I$     & $i\sigma_z$& $ \sigma_x K$&\ 1,\ 1,$B_1$ &2& $(S_{xy},S_{xy},S_{xyz})$ &ladder \\
\hline
$E_3$       & $ \sigma_z$&    $I$     & $i\sigma_y K$&\ 1,-1,$B_3$ &2& $(S_x,S_x,S_{xyz})$ &ladder\\
$E_3'$      & $ \sigma_z$&    $I$     & $i\sigma_x K$&\ 1,\ 1,$B_3$ &2& $(S_{yz},S_{yz},S_{xyz})$& ladder\\
\hline
$E_5$       & $i\sigma_z$& $ \sigma_x$&    $I K$     &-1,\ 1,$A$ &2& $(S_{yz},S_y,S_{xy})$& chain($T_y$ phase) \\
$E_5'$      & $I\otimes i\sigma_z$& $ I\otimes\sigma_x
$&$\sigma_y\otimes I K$      &-1,-1,$A$ &4&$(S_{xyz}^3,
S_x^3,S_{yz}^1, S_{xz}^3,S_y^1,
S_z^3,S_{xy}^1)$\footnote{$(S_{xyz}^3, S_x^3,S_{yz}^1,
S_{xz}^3,S_y^1, S_z^3,S_{xy}^1)=
(S_{xyz},S_{xyz},S_{xyz},S_{x},S_{x},S_{x},S_{yz},S_{xz},S_{xz},S_{xz},S_{y},
S_{z},S_{z},S_{z},S_{xy} )$. Here $S^3_x$, for example, means that
$S_x$ appears for three times: $S^3_x \to S_{x},S_{x},S_{x}$.  Also,
these three $S_{x},S_{x},S_{x}$ do not correspond to the same
physical operator. They correspond to three different operators that
transform in the same way as the $S_x$ operator. For instance, they
may correspond to $S_x$ at three different sites near the end spin.
}  &ladder\\
\hline
$E_7$       & $ \sigma_z$& $i\sigma_z$& $i\sigma_x K$&\ 1,\ 1,$B_2$ &2&$(S_{xz},S_{xz},S_{xyz})$ & ladder\\
$E_7'$      & $ \sigma_z$& $i\sigma_z$& $i\sigma_y K$&\ 1,-1,$B_2$ &2&$(S_y,S_y,S_{xyz})$& ladder\\
\hline
$E_9$       & $i\sigma_z$& $ \sigma_x$& $i\sigma_x K$&-1,\ 1,$B_3$ &2& $(S_{yz},S_{xz},S_z)$ & chain($T_z$ phase)\\
$E_9'$      & $I\otimes i\sigma_z$& $ I\otimes \sigma_x$&$\sigma_y\otimes i\sigma_x K$&-1,-1,$B_3$ &4&$(S_{xyz}^3, S_x^3,S_{yz}^1, S_y^3,S_{xz}^1 ,S_{xy}^3,S_z^1)$\footnote{$(S_{xyz}^3, S_x^3,S_{yz}^1, S_y^3,S_{xz}^1 ,S_{xy}^3,S_z^1)= (S_{xyz},S_{xyz},S_{xyz}, S_{x},S_{x},S_{x},S_{yz},S_{y},S_{y},S_{y},S_{xz}, S_{xy},S_{xy},S_{xy},S_{z} )$.
}  & ladder\\
\hline
$E_{11}$    & $i\sigma_z$& $i\sigma_x$& $ \sigma_z K$&-1,\ 1,$B_1$ &2& $(S_x,S_{xz},S_{xy})$ & chain($T_x$ phase)\\
$E_{11}'$   & $I\otimes i\sigma_z$& $I\otimes i\sigma_x$& $ \sigma_y\otimes \sigma_z K$&-1,-1,$B_1$ &4&$(S_{xyz}^3, S_{yz}^3,S_x^1, S_y^3,S_{xz}^1, S_z^3,S_{xy}^1)$\footnote{$(S_{xyz}^3, S_{yz}^3,S_x^1, S_y^3,S_{xz}^1, S_z^3,S_{xy}^1)= (S_{xyz},S_{xyz},S_{xyz}, S_{yz},S_{yz},S_{yz},S_{x}, S_{y},S_{y},S_{y},S_{xz}, S_{z},S_{z},S_{z},S_{xy} )$.}  & ladder\\
\hline
$E_{13}$    & $i\sigma_z$& $i\sigma_x$& $i\sigma_y K$&-1,-1,$B_2$ &2& $(S_x,S_y,S_z)$ & chain($T_0$ phase)\\
$E_{13}'$   & $I\otimes i\sigma_z$& $I\otimes i\sigma_x$&$\sigma_y\otimes i\sigma_y K$& -1,\ 1,$B_2$ &4&$(S_{xyz}^3, S_{yz}^3,S_x^1, S_{xz}^3,S_y^1, S_{xy}^3,S_z^1)$\footnote{$(S_{xyz}^3, S_{yz}^3,S_x^1, S_{xz}^3,S_y^1, S_{xy}^3,S_z^1)= (S_{xyz},S_{xyz},S_{xyz}, S_{yz},S_{yz},S_{yz},S_{x}, S_{xz},S_{xz},S_{xz},S_{y}, S_{xy},S_{xy},S_{xy},S_{z} ).$}  & ladder\\
\end{tabular}
\end{ruledtabular}
\end{table*}

All the linear representations of the group anti-unitary $D_2+T$ are 1-dimensional (1-D).
The number of linear representations of depends on the representation space.
When acting on Hilbert space, the linear representations are classified by $H^1(D_2+T,U_T(1))=(Z_2)^2$,
which contains four elements. When acting on Hermitian operators, the linear representations
are classified by $H^1(D_2+T,(Z_2)_T)=(Z_2)^3$, which contains eight elements. More details about
linear representations and the first group cohomology are given in appendix \ref{app0}. The 8 linear
representations (with Hermitian operators as the representation space)
are shown in Table~\ref{tab:Repd2h}. These 8 representations collapse into 4 if
the representation space is a Hilbert space, because the bases $|1,x\rangle$
and $i|1,x\rangle$ (similarly, $|1,y\rangle$ and $i|1,y\rangle$,
$|1,z\rangle$ and $i|1,z\rangle$, $|0,0\rangle$ and $i|0,0\rangle$) are not independent.
In the following discussion, if there is no further clarification, we will assume the
linear representations are defined on a Hermitian operator space. Some of these Hermitian 
operators, called active operators which will be defined later, are very important to distinguish different STP phases.


The projective representations are classified
by the group cohomology $H^2(D_2+T,U_T(1))$. There are totally 16
different classes of projective representations for $D_2+T$, as
shown in Tabel~\ref{tab:prjRep2}. More discussions about group
cohomology and projective representation are given in appendices
\ref{app:cohomology}, \ref{app:revclassify}, \ref{app:prjD2+T} and
\ref{app:realizeSPT}. The 16 classes of projective representations
correspond to 16 SPT phases. Our result agrees with the
classification in Ref.~\onlinecite{CGW1123}, and the correspondence
is illustrated by the indices $(\omega(D_2), \beta(T),
\gamma(D_2))$.

In all these 16 SPT phases, the bulk is gapped and we can only
distinguish them by their different edge states which are described
by the projective representations. We stress that all the properties
of each SPT phase are determined by the edge states and can be
detected experimentally. The idea is to add various perturbations
that break the $D_2+T$ symmetry, and to see how those perturbations
split the degeneracy of the edge states.

Let us firstly consider the case that the space of degenerate end
`spin' is 2-dimensional. We have three Pauli matrices $(\sigma_x,
\sigma_y, \sigma_z)$ to lift the end `spin' degeneracy. During
various perturbations of the system, only those that reduce to the
Pauli matrices $(\sigma_x, \sigma_y, \sigma_z)$ can split the
degeneracy of the ground states. These perturbations will be called
active operators. To identify whether a perturbation is an active
operator, one can compare its symmetry transformation properties
under $D_2+T$ with those of the three Pauli matrices
$(\sigma_x,\sigma_y,\sigma_z)$.
For different SPT phases, the end spin forms different projective
representations of the $D_2+T$ group, and consequently the three
Pauli matrices $(\sigma_x,\sigma_y,\sigma_z)$ form different linear
representations of $D_2+T$. So they correspond to different active
operators in different SPT phases.

Let $O$ be a perturbation operator, under the symmetry
operation $g$ it varies as
\begin{eqnarray}
u(g)^{\dag} O u(g)&=&\eta_g(O) O,
\end{eqnarray}
where $u(g)$ is the representation of symmetry transformation $g$ on
the physical spin Hilbert space, $\eta_{g}(O)$ is equal to 1 or -1
and forms a 1-D representation of the symmetry group $D_2+T$. On the
other hand, the three Pauli matrices $(\sigma_x, \sigma_y,
\sigma_z)$ also form linear representations of $D_2+T$. In the end
`spin' space, the Pauli matrices transform as ($m=x,y,z$)
\begin{eqnarray}
M(g)^\dag \sigma_m M(g)&=&\eta_g(\sigma_m) \sigma_m,
\end{eqnarray}
where $M(g)$ is the projective representation of $g$ (see
Tabel~\ref{tab:prjRep2}) on the end `spin' Hilbert space. If the
physical operator $O$ and the end `spin' operator $\sigma_m$ form
the same linear representation of the symmetry group, namely,
$\eta_g(O)=\eta_g(\sigma_m)$, then they should have the same matrix
elements (up to a constant factor) in the end spin subspace. In
Table~\ref{tab:prjRep2}, the sequence of operators $(O_1,O_2,O_3)$
are the active operators corresponding to the end `spin' operators
$(\sigma_x, \sigma_y, \sigma_z)$, respectively.

Similarly, in the case that the end `spin' is 4-dimensional, there
are 15 $4\times4$ matrices that can (partially) lift the degeneracy
of the end states, namely, $(\si_x\otimes I,\si_y\otimes
I,\si_z\otimes I,
\si_x\otimes\si_x,\si_y\otimes\si_x,\si_z\otimes\si_x,I\otimes\si_x,
\si_x\otimes\si_y,\si_y\otimes\si_y,\si_z\otimes\si_y,I\otimes\si_y,
\si_x\otimes\si_z,\si_y\otimes\si_z,\si_z\otimes\si_z,I\otimes\si_z)$.
And the corresponding active operators are given in
Table~\ref{tab:prjRep2}.

Since the active operators are perturbations that spilt the ground
state degeneracy, through linear response theory, they correspond to
measurable physical quantities. For example, if the spin $S_m$ is an
active operator, it couples to a magnetic field through the
interaction
\begin{eqnarray}\label{B}
H'=\sum_i\left(g_x\mu_BB_xS_{x,i} + g_y\mu_BB_yS_{y,i} +
g_z\mu_BB_zS_{z,i}\right).
\end{eqnarray}
The end `spins' may be polarized by above perturbation. In a real
spin-chain materials, due to structural defects, there are
considerable number of end `spins'. They behave as impurity spins
(the gapped bulk can be seen as a paramagnetic material). Thus, the
polarizing of the end `spins' can be observed by measuring the
magnetic susceptibility, which obeys the Curie law ($m=x,y,z$)
\begin{eqnarray*}
\chi_m(T)= {Ng_m^2\mu_B\over 3k_BT},
\end{eqnarray*}
where $N$ is the number of end `spins'.

Notice that different projective representations have different
active operators. Thus we can distinguish all of the 16 SPT phases
experimentally. For instance, the active operators of the $E_1$ and
$E_1'$ phases are $(S_z,S_z,S_{xyz})$ and $(S_{xy},S_{xy},S_{xyz})$,
respectively. Here $S_{mn}=S_mS_n+S_nS_m$ is a spin quadrupole
operator, and $S_{xyz}$ is a third order spin operator, such as
$S_{xy,i}S_{zi+1}$ or $S_{x,i}S_{y,i+1}S_{z,i+2}$. We will show that
the two SPT phases $E_1$ and $E'_1$ can be distinguished by the
perturbation (\ref{B}).  In $E_1$ phase, the active operators
contain $S_z$, so it response to $B_z$. In consequence, the
$g$-factors $g_z$ is finite, but $g_x, g_y = 0$ (because $S_x, S_y$
are not active operators). However, in $E_1'$ phase, none of $S_x,
S_y, S_z$ is active, so the end `spins' do not response to magnetic
field at all. As a consequence, all components of the $g$-factor
approaches zero: $g_{x}, g_y, g_z= 0$. This difference distinguishes
the two phases.

To completely separate all the 16 SPT phases, one need to add
perturbations by the spin-quadrupole operators $S_{xy}, S_{yz},
S_{xz}$ and the third-order spin operators such as
$S_{xy,i}S_{z,i+1}$. Actually, these perturbations may be realized
experimentally. For instance, the interaction between the
spin-quadrupole and a nonuniform magnetic field is reasonable in
principle:
\[
H'=g_{xy}\left(\frac{\partial B_x}{\partial y} + \frac{\partial
B_y}{\partial x}\right)S_{xy}+...
\]
One can measure the corresponding `quadrupole susceptibility'
corresponding to above perturbation. Similar to the spin
susceptibility, different SPT phases have different coupling
constants for the `quadrupole susceptibility'. Consequently, from
the information of the spin dipole- and quadrupole- susceptibilities
(and other information corresponding to the third-order spin
operators), all the 16 SPT phases can be distinguished.

\section{Realization of SPT phases in $S=1$ spin chains and
ladders}\label{sec. realizations}

In this section, we will illustrate that all these 16 SPT
phases can be realized in $S=1$ spin chains or ladders.

\subsection{spin-chains }\label{sec.chain}
\subsubsection{SPT phases for nontrivial projective representations}\label{sec.triSPT}

In Ref.~\onlinecite{LLW1180}, we have studied four nontrivial SPT
phases $T_0, T_x, T_y, T_z$ in $S=1$ spin chains. The ground states
of these phases are written as a matrix product state (MPS)
\begin{eqnarray*}
|\phi\rangle=\sum_{\{m_i\}}\mathrm{Tr}(A^{m_1}_1A^{m_2}_2...A^{m_N}_N)|m_1m_2...m_N\rangle.
\end{eqnarray*}
where $m_i=x,y,z$. More information about MPS is given in
appendix.~\ref{app:revclassify}.

1) $T_0$ phase. The end `spins' of this phase belong to the
projective representation $E_{13}$, and a typical MPS in this phase
is
\begin{eqnarray}
A^{x}=a\sigma_x,\ \ \
A^{y}=b\sigma_y,\ \ \
A^{z}=c\sigma_z,
\end{eqnarray}
where $a,b,c$ are real numbers.\cite{note:SO(3)} Table~\ref
{tab:prjRep2} shows that the active operators in this phase are
$S_x,  S_y,  S_z $, so the end spins will response to the magnetic
field along all the three directions.

2) $T_x$ phase. The end `spins' of this phase belong to the
projective representation $E_{11}$, and a typical MPS in this phase
is
\begin{eqnarray}
A^{x}=a\sigma_x, A^{y}=ib\sigma_y, A^{z}=ic\sigma_z,
\end{eqnarray}
where $a,b,c$ are real numbers. Table~\ref{tab:prjRep2} shows that
there is only one active operator $S_x$ in this phase, so the end
spins will only response to the magnetic field along $x$ direction.

3) $T_y$ phase. The end `spins' of this phase belong to the
projective representation $E_{5}$, and a typical MPS in this phase
is
\begin{eqnarray}
A^{x}=ia\sigma_x, A^{y}=b\sigma_y, A^{x}=ic\sigma_z,
\end{eqnarray}
where $a,b,c$ are real numbers. Table~\ref{tab:prjRep2} shows that
there is only one active operator $S_y$ in this phase, so the end
spins will only response to the magnetic field along $y$ direction.

4) $T_z$ phase. The end `spins' of this phase belong to the
projective representation $E_{9}$, and a typical MPS in this phase
is
\begin{eqnarray}
A^{x}=ia\sigma_x, A^{y}=ib\sigma_y, A^{z}=c\sigma_z,
\end{eqnarray}
where $a,b,c$ are real numbers. Table~\ref{tab:prjRep2} shows that
there is only one active operator $S_z$ in this phase, so the end
spins will only response to the magnetic field along $z$ direction.

\subsubsection{SPT phases for trivial projective representations}

Corresponding to the trivial projective IRs, we can also construct
trivial phases. Here `trivial' means that the ground state is in
some sense like a direct product state. In these phase the matrix
$A^m$ also vary as Eqs.~(\ref{MAM}) and (\ref{MAMT}), except that
$A^m$ is a 1-D matrix, and $M(g)$ is a 1-d representation
of $D_2+T$.
Since all the 1-D representation belongs to the same
class, there is only one trivial phase.

A simple example of the states in this phase is a
direct product state
\[|\phi\rangle=|m\rangle_1|m\rangle_2...|m\rangle_N.\]
This state can be realized by a strong (positive) on-site single-ion
anisotropy term $(S_m)^2$, $m=x,y,z$. In this phase, there is no
edge state, and no linear response to all perturbations.

%
%


\subsection{spin ladders}

In last section we have realized 5 of the 16 different SPT phases
(with only $D_2+T$ symmetry) in $S=1$ spin chains. In this section,
we will show that all the other phases can be realized in $S=1$
ladders.

\subsubsection{General discussion for spin ladders}\label{sec.ladder.gen}
For simplicity, we will consider the spin-ladder models without
inter-chain interaction.\cite{laddernote} In that case, the ground
state of the spin ladder is a direct product of the ground states of
the independent chains. For example, for a two-leg ladder, the
physical Hilbert space at each site is a direct product space
$\mathcal H=\mathcal H_1\otimes \mathcal H_2$ spanned by bases
$|m_1n_1\rangle=|m_1\rangle|n_1\rangle$, with $m_1,n_1=x,y,z$. If
the ground state of the two chains are $|\phi_1\rangle$ and
$|\phi_2\rangle$ respectively,
\begin{eqnarray}
&&|\phi_1\rangle=\sum_{\{m\}}\mathrm{Tr}(A^{m_1}...A^{m_N})|m_1...m_N\rangle,\nonumber\\
&&|\phi_2\rangle=\sum_{\{n\}}\mathrm{Tr}(B^{n_1}...B^{n_N})|n_1...n_N\rangle,
\end{eqnarray}
with
\begin{eqnarray}
&&\sum_{m'}u(g)_{mm'}A^{m'}=e^{i\alpha_1(g)}M(g)^\dag A^mM(g),\nonumber\\
&&\sum_{n'}v(g)_{nn'}B^{n'}=e^{i\alpha_2(g)}N(g)^\dag B^nN(g),
\end{eqnarray}
for an unitary operator $\hat g$ and
\begin{eqnarray}
&&\sum_{m'}u(T)_{mm'}(A^{m'})^*=M(T)^\dag A^mM(T),\nonumber\\
&&\sum_{n'}v(T)_{nn'}(B^{n'})^*=N(T)^\dag B^nN(T),
\end{eqnarray}
for the time reversal operator $T$. Then the ground state of the
ladder is
\begin{eqnarray}
|\phi\rangle&=&|\phi_1\rangle\otimes|\phi_2\rangle\nonumber
\\&=&\sum_{\{m,n\}}\mathrm{Tr}(A^{m_1}...A^{m_N}) \mathrm{Tr}(B^{n_1}...B^{n_N})|m_1n_1...m_Nn_N\rangle\nonumber\\
&=&\sum_{\{m,n\}}\mathrm{Tr}[(A^{m_1}\otimes
B^{n_1})...(A^{m_N}\otimes B^{n_N})]\nonumber\\&&\ \ \ \ \ \
\times|m_1n_1...m_Nn_N\rangle
\end{eqnarray}
which satisfies
\begin{eqnarray}\label{MN}
&& \sum_{m,n,m',n'}[u(g)\otimes v(g)]_{mn,m'n'}(A^{m'}\otimes
B^{n'})\nonumber\\&& =e^{i\alpha(g)}(M\otimes N)^\dag (A^m\otimes
B^n) (M\otimes N)
\end{eqnarray}
for an unitary $\hat g$ (here $\alpha(g)=\alpha_1(g)+\alpha_2(g)$)
and
\begin{eqnarray}\label{MNT}
&& \sum_{m,n,m',n'}[u(T)\otimes v(T)]_{mn,m'n'}(A^{m'}\otimes
B^{n'})^*\nonumber\\&& =(M\otimes N)^\dag (A^m\otimes B^n) (M\otimes
N)
\end{eqnarray}
for the time reversal operator $T$. This shows that the ground state
of the ladder is also a MPS which is represented by $A^m\otimes
B^n$, and $M\otimes N$ is a projective representation of the
symmetry group $G$.

Specially, if $B^n$ is 1-D and $N(g)=1$ (representing a
trivial phase),  then we have
\begin{eqnarray}\label{Malpha}
&& \sum_{m,n,m',n'}[u(g)\otimes v(g)]_{mn,m'n'}(A^{m'}\otimes
B^{n'})\nonumber\\&&=e^{i\alpha(g)}M^\dag (A^m\otimes B^n)M.
\end{eqnarray}
In general the projective representation $M(g)\otimes N(g)$ is
reducible. This means that the end `spin' of the ladder is a direct
sum space of several irreducible projective representations (IPRs).
These IPRs are degenerate and belong to the same class. However,
this degeneracy is accidental, because only irreducible
representation protected by symmetry is robust. Notice that we
didn't consider the inter-chain interaction in the ladder. If
certain interaction is considered, the degeneracy between the same
classes of IPRs can be lifted, and only one IPR remains as the end
`spin' in the ground state. This IPR (or more precisely the class it
belongs to) determines which phase the spin ladder belongs to.

\begin{table*}[htbp]
\caption{Projective representations of group $\bar
D_2=\{E,R_zT,R_xT,R_y\}$. There are 4 classes of projective
representations, meaning that the second group cohomology contains 4
elements.} \label{tab:D2bar_proj}
\begin{ruledtabular}
\begin{tabular}{c|cccc|c|c|c}
class&E&$R_y$&$R_zT$&$R_xT$& dimension &effecive/active operators & spin models ($S=1$)\\
\hline
1  & 1 &     $  1$  &   $  K$   &    $  K$  &1& &chain(trivial phase)\\
   &$I$&    $\sigma_y$     & $ i\sigma_z K$& $ \sigma_x K$&2& $\sigma_x\sim S_z, S_{yz};\ \sigma_y\sim S_{y};\ \sigma_z\sim  S_x, S_{xy}$& chain \\
\hline
2  &$I$&     $  I$ &   $  \sigma_y K$  &    $  \sigma_y K$ &2&$\sigma_x,\sigma_y,\sigma_z\sim S_{xz}$&ladder\\
   &$I\otimes I$ &    $I\otimes\sigma_y$& $\sigma_y\otimes i\sigma_z K$& $ \sigma_y\otimes\sigma_x K$      &4&  &ladder\\
\hline
3   &$I$& $i\sigma_z$ &    $\sigma_y K$     & $\sigma_x K$&2& $\sigma_x,\sigma_y\sim S_z,S_{yz};\ \sigma_z\sim S_{xz}$ &chain \\
    &$I$& $\sigma_y $ &    $i\sigma_yK$     &      $iIK$  &2& $\sigma_x,\sigma_z\sim S_z,S_{yz};\ \sigma_y\sim S_{xz}$ &chain \\%
\hline
4   &$I$& $i\sigma_z$ &    $\sigma_x K$     & $\sigma_y K$&2& $\sigma_x,\sigma_y\sim S_x, S_{xy};\ \sigma_z\sim S_{xz}$ &chain  \\
    &$I$& $\sigma_y$  &    $iIK       $     & $i\sigma_yK$&2&$\sigma_x,\sigma_z\sim S_x, S_{xy};\ \sigma_y\sim S_{xz}$ &chain %
\end{tabular}
\end{ruledtabular}
\end{table*}

\begin{table*}[htbp]
\caption{Projective representations of group
$Z_2+T=\{E,R_z,T,R_zT\}$.} \label{tab:Z2+T_proj}
\begin{ruledtabular}
\begin{tabular}{c|cccc|c|c|c}
class&E&$R_z$&$T$&$R_zT$& dimension &effecitive/active operators &spin models ($S=1$)\\
\hline
1   & 1 &   $  1$   &    $  K$  &     $  K$  &1& &chain\\
    &$I$&    $\sigma_y$     & $ i\sigma_z K$& $ \sigma_x K$&2& $\sigma_x\sim  S_x,S_y;\ \sigma_y\sim S_{xy};\ \sigma_z\sim  S_{yz}, S_{xz} $& chain \\
\hline
2   &$I$&   $  I$   &    $  \sigma_y K$ &     $  \sigma_y K$ &2&$\sigma_x,\sigma_y,\sigma_z\sim S_{z}$&ladder\\
    &$I\otimes I$ &    $I\otimes\sigma_y$& $\sigma_y\otimes i\sigma_z K$& $ \sigma_y\otimes\sigma_x K$      &4&  &ladder\\
\hline
3   &$I$&    $i\sigma_z$     & $\sigma_y K$& $\sigma_xK$ &2& $\sigma_x,\sigma_y\sim S_x, S_{y};\ \sigma_z\sim S_{z}$ &chain \\
    &$I$& $\sigma_y $ &    $i\sigma_yK$     &      $iIK$  &2& $\sigma_x,\sigma_z\sim S_x, S_{y};\ \sigma_y\sim S_{z}$  &chain \\%
\hline
4   &$I$&    $i\sigma_z$     & $\sigma_x K$ & $\sigma_yK$ &2& $\sigma_x,\sigma_y\sim S_{xz}, S_{yz};\ \sigma_z\sim S_{z}$ &chain  \\
    &$I$& $\sigma_y$  &    $iIK       $     & $i\sigma_yK$&2&$\sigma_x,\sigma_z\sim S_{xz}, S_{yz};\ \sigma_y\sim S_{z}$  &chain %
\end{tabular}
\end{ruledtabular}
\end{table*}

\subsubsection{$S=1$ spin ladders in different SPT phases}

In appendix \ref{app:realizeSPT}, we show how to obtain all the
other IPRs by reducing the direct product representations of
$E_{13}, E_{11}, E_5, E_9$. We start with these four IPRs because
the corresponding SPT phases $T_0,T_x,T_y,T_z$ have been realized in
spin chains. Actually, the reduction procedure provides a method to
construct spin ladders from spin chains and to realize all the SPT
phases.

By putting two different spin chains (belonging to the
$T_0,T_x,T_y,T_z$ phases) into a ladder, we obtain 6 new phases
corresponding to $E_1, E_1', E_3, E_3', E_7, E_7'$, respectively. If
we put one more spin chain into the ladder, then we obtain 5 more
new phases corresponding to $E_0', E_5', E_9', E_{11}', E_{13}'$,
respectively. Therefore, together with $T_0,T_x,T_y,T_z$ and the
trivial phase in spin chains, we have realized all the 16 SPT phases
listed in Table~\ref{tab:prjRep2}. Furthermore, if we have
translational symmetry, then from section \ref{sec.triSPT} and
Eq.~(\ref{Malpha}), we have totally $16\times4=64$ different SPT
phases in spin ladders, in accordance with the result of
Ref.~\onlinecite{CGW1107}.

\section{SPT phases for subgroups of $D_{2}+T$}\label{sec. sugroups}
From the projective representations of group $D_2+T$, we can easily
obtain the projective representations of its subgroups. According to
Table~\ref{tab:prjRep2}, the representation matrices for the
subgroups also form a projective representation, but usually it is
reducible. By reducing these matrices, we can obtain all the IPRs of
the subgroup.

\subsection {$\bar D_2=\{E,R_zT,R_xT,R_y\}$}
This group is also a $D_2$ group except that half of its elements
are anti-unitary. Notice that $T$ itself is not a group element.
This group has four 1-D linear representations. In
Table~\ref{tab:D2bar_linear} in appendix \ref{app0}, we list the
representation matrix elements, representational bases of physical
spin and spin operators (for $S=1$) according to each linear
representation.

The projective representations of the subgroup $\bar D_2$ are shown
in Table~\ref{tab:D2bar_proj}. By reducing the representation matrix
of $D_2+T$, we obtained 8 projective representations. They are
classified into 4 classes. This can be shown by calculating the
corresponding 2-cocycles of these projective representations. Two
projective representations belonging to the same class means that
the corresponding 2-cocycle differ by a 2-coboundary (see appendices
\ref{app:cohomology}, \ref{app:revclassify} and \ref{app:prjD2+T}).


As shown in Table~\ref{tab:D2bar_proj}, the 2-dimensional
representation in class-1 is trivial (or linear), it belongs to the
same class as the 1-D representation. This means that the
edge states in this phase is not protected by symmetry, the ground
state degeneracy can be smoothly lifted without phase transition.
The class-3 and class-4 nontrivial SPT phases can be realized in
spin chains. These two phases can be distinguished by magnetic
fields.
The phase corresponding to the class-3 
projective representation only response to the magnetic field along
$z$ direction, and the phase corresponding to class-4 
projective representation only respond to the magnetic field along
$x$ direction.
The remaining two nontrivial SPT phases of class 2 can be realized
by spin ladders. 


\subsection{$Z_2+T=\{E,R_z,T,R_zT\}$}

This subgroup is also a direct product group. The linear
representations and projective representations are given in
Table.~\ref{tab:Z2+T_linear} (see appendix \ref{app0}) and
\ref{tab:Z2+T_proj}, respectively. This group is isomorphic to $\bar
D_2=\{E,R_zT,R_xT,R_y\}$ , so its projective representations and SPT
phases are one to one corresponding to those in
\ref{tab:D2bar_proj}. However, the corresponding SPT phases in
\ref{tab:Z2+T_proj} and \ref{tab:D2bar_proj} are not the same,
because they have different response to external perturbations.


Notice that, this simple symmetry is very realistic for materials.
For example, the quasi-1D anti-ferromagnets CaRuO$_3$\cite{CaRuO3}
and NaIrO$_3$ \cite{NaIrO3} respect this $Z_2+T$ symmetry due to
spin-orbital coupling. Their ground state, if non-symmetry breaking,
should belong to one of the four SPT phases listed in
Table~\ref{tab:Z2+T_proj}.


\section{conclusion and discussion}\label{sec. summary}

In summary, through the projective representations, we studied all
the 16 different SPT phases for integer spin systems that respect
only $D_{2h}=D_2+T$ on-site symmetry. We provided a method to
measure all the SPT orders. We showed that in different SPT phase
the end `spins' respond to various perturbations differently. These
perturbations include spin dipole- (coupling to uniform magnetic
fields) and quadrupole- operators (coupling to nonuniform magnetic
fields). We illustrated that the SPT orders in different SPT phases
can be observed by experimental measurements, such as the
temperature dependence of the magnetic susceptibility and asymmetric
$g$-factors. We illustrated that all the 16 SPT phases can be
realized in $S=1$ spin chains or ladders. Finally we studied the SPT
phases for two subgroups of $D_2+T$, one of the subgroup is the
symmetry group of some interesting materials.\cite{CaRuO3,NaIrO3}
Certainly, our method of studying SPT orders can be generalized to
other symmetry groups.

\section{acknowledgements}
We thank Ying Ran for helpful discussions. This research is
supported by NSF Grant No. DMR-1005541 and NSFC 11074140.

\appendix

\section{Group cohomology}\label{app:cohomology}

We consider a finite group $G=\{g_1,g_2,...\}$ with its module space
$U_T(1)$. The group elements of $G$ are operators on the module space.
A \textit{$n$-cochain} $\omega_n(g_1,g_2,...,g_n)$ is a function on
the group space which maps $\otimes^nG\to U(1)$. The cochains can be
classified with the \textit{coboundary operator}.

Suppose the cochain $\omega_n(g_1,g_2,...,g_n)\in U(1)$, then the
coboundary operator is defined as
\begin{eqnarray*}
&&(d\omega_n)(g_1,g_2,...,g_{n+1})=g_1\cdot\omega_n(g_2,g_3,...,g_{n+1})\\
&& \omega_n^{-1}(g_1g_2,g_3,...,g_{n+1})\ \omega_n(g_1,g_2g_3,...,g_{n+1})\ ...\\
&& \omega_n^{(-1)^i}(g_1,g_2,...,g_ig_{i+1},...,g_{n+1})\ ...\\&&
\omega_n^{(-1)^{n}} (g_1,g_2,...,g_{n}g_{n+1})
\omega_n^{(-1)^{n+1}}(g_1,g_2,...,g_{n}),
\end{eqnarray*}
for $n\geq1$, and
\begin{eqnarray}\label{CB1}
&&(d\omega_0)(g_1)=\frac{g_1\cdot\omega_0}{\omega_0},
\end{eqnarray}
for $n=0$. Here $g\cdot\omega_n$ is a group action on the module space $U(1)$.
If $g$ is an unitary operator, it acts on $U(1)$ trivially $g\cdot\omega_n
=\omega_n$. If $g$ is anti-unitary (such as the time reversal operator $T$),
then the action is given as $g\cdot\omega_n= \omega_n^*=\omega_n^{-1}$.  We
will use $U_T(1)$ to denote such a module space.  We note that, if $G$ contain
no time reversal transformation, then $U_T(1)=U(1)$.

A cochain $\omega_n$ satisfying $d\omega_n=1$ is called a \textit{
$n$-cocycle}. If $\omega_n$ satisfies $\omega_n=d\omega_{n-1}$, then
it is called a \textit{$n$-coboundary}. Since $d^2\omega=1$, a
coboundary is always a cocycle. The following are two examples of
cocycle equations. 1-cocycle equation:
\begin{eqnarray}\label{1Cocyc}
\frac{g_1\cdot\omega_2(g_2)\omega(g_1)}{\omega_2(g_1g_2)}=1.
\end{eqnarray}
2-cocyle equation:
\begin{eqnarray}\label{2Cocyc}
\frac{g_1\cdot\omega_2(g_2,g_3)\omega_2(g_1,g_2g_3)}{\omega_2(g_1g_2,g_3)\omega_2(g_1,g_2)}=1.\nonumber\\
\end{eqnarray}

The \textit{group cohomology} is defined as $H^n(G,U_T(1))=Z^n/B^n$.
Here $Z^n$ is the set of $n$-cocycles and $B^n$ is the set of
$n$-coboundarys. If two $n$-cocycles $\omega_n$ and $\omega'_n$
differ by a  $n$-coboundary $\tilde\omega_n$, namely, $\omega'_n=
\omega_n\tilde\omega_n^{-1}$, then they are considered to be
equivalent. The set of equivalent $n$-cocycles is called a
equivalent class. Thus, the $n$-cocycles are classified with
different equivalent classes, these classes form the (Abelian)
cohomology group $H^n(G,U_T(1))=Z^n/B^n$.

As an example, we see the cohomology of $Z_2=\{E,\sigma\}$, where
$E$ is the identity element and $\sigma^2=E$. Since this group $Z_2$
is unitary, it acts on the module space trivially and $U_T(1)=U(1)$:
$g\cdot\omega_n=\omega_n$. From (\ref{1Cocyc}) the first cohomology
is the 1-D representations.
\[
H^1(Z_2,U(1))=Z_2,
\]
The second cohomology classifies the projective representations (see
appendix \ref{app:revclassify}). It can be shown that all the
solutions of (\ref{2Cocyc}) are 2-coboundaries $\omega_2=d\omega_1$.
So all the 2-cocycles belong to the same class, consequently,
\[
H^2(Z_2,U(1))=0.
\]

Let us see another example, the time reversal group $Z_2^T=\{E,T\}$.
Notice that the time reversal operator $T$ is antiunitary, it acts on
$U_T(1)$ nontrivially: $T\cdot\omega_n=\omega_n^{-1}$. As a result,
the cohomology of $Z_2^T$ is different from that of $Z_2$:
\begin{eqnarray*}
&&H^1(Z_2^T,U_T(1))=0,\\
&&H^2(Z_2^T,U_T(1))=Z_2.
\end{eqnarray*}
The group  $Z_2^T$ have two orthogonal 1-d representations (see
appendix \ref{app0}), but above result shows that these two
1-D representations belongs to the same class. Further
more, the nontrivial second group cohomology shows that $Z_2^T$ has
a nontrivial projective representation, which is well known:
$M(E)=I, M(T)=i\sigma_yK$.

\section{Brief review of the classification of 1D SPT
orders}\label{app:revclassify}

A key trick to use local unitary transformation to study/classify 1D
gapped SPT phases is the matrix product state (MPS) representation
of the ground states. The simplest example is the $S=1$ AKLT wave
function \cite{AKL8877} in the Haldane phase which can be written as
a $2\times2$ MPS. Later it was shown that in 1D all gapped many-body
spin wave functions (it was generalized to fermion systems) can be
well approximated by a MPS as long as the dimension $D$ of the
matrix is large enough \cite{MPS}
\begin{eqnarray}\label{MPS}
|\phi\rangle=\sum_{\{m_i\}}\mathrm{Tr}(A^{m_1}_1A^{m_2}_2...A^{m_N}_N)|m_1m_2...m_N\rangle.
\end{eqnarray}
Here $m$ is the index of the $d$-component physical spin, and
$A^{m_i}_i$ is a $D\times D$ matrix. Provided that the system is
translationally invariant, then one set all the matrices $A^m$ as
the same over all sites.

In the MPS picture, it is natural to understand that projective
representations can be used as a label of different SPT phase.
Suppose that a system has an on-site unitary symmetry group $G$
which keep the ground state $|\phi\rangle$ invariant
\begin{eqnarray}\label{g|psi>}
\hat g|\phi\rangle= u(g)\otimes u(g)\otimes...\otimes u(g) |\phi
\rangle=(e^{i\alpha(g)})^N|\phi\rangle,
\end{eqnarray}
where $\hat g\in G$ is a group element of $G$, $u(g)$ is its
$d$-dimensional (maybe reducible) representation and
$e^{i\alpha(g)}$ is its 1-D representation. We only
consider the case that $u(g)$ is a linear presentation of $G$. The
case that $u(g)$ forms a projective representation of $G$ (such as
half-integer spin chain) has been studied in \Ref{CGW1107,CGW1123}.
Eqs.~(\ref{MPS}) and (\ref{g|psi>}) require that the matrix $A^m$
should vary in the following way\cite{PBT1039,CGW1107}
\begin{eqnarray}\label{MAM}
\sum_{m'}u(g)_{mm'}A^{m'}=e^{i\alpha(g)}M(g)^\dag A^m M(g),
\end{eqnarray}
where $M(g)$ is an invertible matrix and is essential for the
classification of different SPT phases. Notice that if $M(g)$
satisfies Eq.~(\ref{MAM}), so does $M(g)e^{i\varphi(g)}$. Since
$u(g_1g_2)=u(g_1)u(g_2)$ and $e^{i\alpha(g_1g_2)}=e^{i\alpha(g_1)}
e^{i\alpha(g_2)}$, we obtain
\begin{eqnarray}
M(g_1g_2)=M(g_1)M(g_2)e^{i\theta(g_1,g_2)}.
\end{eqnarray}
Above equation shows that up to a phase $e^{i\theta(g_1,g_2)}$,
$M(g)$ satisfies the multiplication rule of the group. Further,
$M(g)$ satisfies the associativity condition
$M(g_1g_2g_3)=M(g_1g_2)M(g_3)e^{i\theta(g_1g_2,g_3)}=M(g_1)
M(g_2g_3)e^{i\theta(g_1,g_2g_3)}$, or equivalently
\[
e^{i\theta(g_2,g_3)}e^{i\theta(g_1,g_2g_3)}
=e^{i\theta(g_1,g_2)}e^{i\theta(g_1g_2,g_3)} .
\]
Above equation coincide with the cocycle equation (\ref{2Cocyc})
when $G$ is unitary. The matrices $M(g)$ that satisfies above
conditions are called \textit{projective representation} of the
symmetry group $G$. Above we also shows the relation between
projective representations and 2-cocycle.

For a projective representation, the two-element function
$e^{i\theta(g_1,g_2)}$ has redundant degrees of freedom. Suppose
that we introduce a phase transformation,
$M(g_1)'=e^{i\varphi(g_1)}M(g_1) $, $M(g_2)'=e^{i\varphi(g_2)}M(g_2)
$ and $M(g_1g_2)'=e^{i\varphi(g_1g_2)}M(g_1g_2)$, then the function
$e^{i\theta(g_1,g_2)}$ becomes
\begin{eqnarray}
e^{i\theta(g_1,g_2)'}=\frac{e^{i\varphi(g_1g_2)}}
{e^{i\varphi(g_1)}e^{i\varphi(g_2)}} e^{i\theta(g_1,g_2)}.
\end{eqnarray}
Notice that $e^{i\theta(g_1,g_2)'}$ and $e^{i\theta(g_1,g_2)}$
differs by a 2-coboundary, so they belong to the same class.
Thus, the projective representations are classified by the second
group cohomology $H^2(G,U_T(1))$. If $M(g)$ and $\tilde M(g)$ belong to different (classes of)
projective representations, then they cannot be smoothly transformed
into each other, therefore the corresponding quantum states $A^m$
and $\tilde A^{m}$ fall in different phases. In other words, the
projective representation $\omega_2\in H^2(G,U_T(1))$ provides a label
of a SPT phase. If the system is translationally invariant, then
$e^{i\alpha(g)}\in H^1(G,U_T(1))$ is also a label of a SPT phase. In
this case, the complete label of a SPT phase is $(\omega_1,\alpha)$.
If translational symmetry is absent, we can regroup the matrix $A^m$
such that $e^{i\alpha(g)}=1$, then each SPT phase is uniquely
labeled by $\omega_2$.

\section{Linear representations for $D_2+T$ and its
subgroups}\label{app0}

Generally, the 1-D linear representations of a group $G$ are classified by its first group cohomology $H^1(G)$. However, there is a subtlety to choose the coefficient of $H^1(G)$. We will show that if the representation space is a Hilbert space, the 1-D representations are characterized by $H^1(G, U(1))$ (or $H^1(G,U_T(1))$ if $G$ contains anti-unitary elements); while if the representation space is a Hermitian operator space, then the 1-D representations are characterized by $H^1(G,Z_2)$ (notice that $H^1(G,(Z_2)_T)=H^1(G,Z_2)$, there is no difference whether $G$ contains anti-unitary elements or not).

Since the discusses for unitary group and anti-unitary group are very similar, we will only consider a group $G$ which contains anti-unitary elements. Firstly, we consider the 1-D linear representations on a Hilbert space $\mathcal H$. Suppose $\phi\in \mathcal H$ is a basis, and $g\in G$ is an anti-unitary element, then
\begin{eqnarray}
\hat g|\phi\rangle=\eta(g)K|\phi\rangle,
\end{eqnarray}
where the number $\eta(g)$ is the representation of $g$. Notice that $g$ is anti-linear, which may change the phase of $|\phi\rangle$. To see that, we suppose $K|\phi\rangle=|\phi\rangle$,  and introduce a phase transformation for the basis $|\phi\rangle$, namely, $|\phi'\rangle=|\phi\rangle e^{i\theta}$. Now we choose $|\phi'\rangle$ as the basis, then
\begin{eqnarray}
\hat g|\phi'\rangle=\eta(g)e^{i2\theta}K|\phi'\rangle,
\end{eqnarray}
so the representation $\eta(g)'=\eta(g)e^{i2\theta}$ changes accordingly. This means that the 1-D representation of the group $G$ is $U(1)$-valued, and is characterized by the first cohomology group $H^1(G,U(1))$. In the case of $D_2+T$, we have \[H^1(D_2+T,U_T(1))=(Z_2)^2,\] so $D_2+T$ has 4 different 1-D linear representations on Hilbert space, which can be labeled as
$A, B_1, B_2, B_3$ respectively. 

Now we consider the 1-D representations on a Hermitian operator space. Suppose $O_1, O_2,...,O_N$ are orthonormal Hermitian operators satisfying $\mathrm {Tr} (O_mO_n)=\delta_{mn}$, an anti-unitary element $g\in G$ act on these operators as
\begin{eqnarray}\label{Omn}
\hat g O_m = KM(g)^\dag O_m M(g)K = \sum_n\zeta(g)_{mn} O_n,
\end{eqnarray}
Here $M(g)K$ is either a linear or a projective representation of $g$, while $\zeta(g)$ is always a linear representation. Since $[KM(g)^\dag O_m M(g)K]^\dag =KM(g)^\dag O_m M(g)K$, we have $[\sum_n\zeta(g)_{mn} O_n]^\dag=\sum_n\zeta(g)_{mn}^* O_n=\sum_n \zeta(g)_{mn} O_n$, which gives \[\zeta(g)^*=\zeta(g).\] The same result can be obtained if $G$ is unitary.  So we conclude that, \textit{all the linear representations defined on Hermitian operator space are real.} Now we focus on 1-D linear representations. Since $g$ is either unitary or anti-unitary, we have $|\zeta(g)|=1$. On the other hand, $\zeta(g)$ must be real, so $\zeta(g)=\pm1$. As a result, all the 1-D linear representations on Hermitian operator space are $Z_2$ valued, which are characterized by the first group cohomology $H^1(G,(Z_2)_T)$. For the group $D_2+T$,
\[
H^1(D_2+T,(Z_2)_T)=(Z_2)^3,
\]
so there are 8 different 1-D linear representation, corresponding to 8 classes of Hermitian operators as shown
in Tabel~\ref{tab:Repd2h}. Since all the linear representations of $D_2+T$ are 1-dimensional, this 8 1-D representations are all of its linear representations.

Above discussion is also valid for the subgroups of $D_2+T$. In Tabels~\ref{tab:D2bar_linear} and \ref{tab:Z2+T_linear}, we give the linear representations of its two subgroups (the number of 1-D linear representations on Hilbert space is half of that on Hermitian operator space).

We have shown that for 1-D linear representations defined on Hermitian operator space, there is no difference whether a group element is unitary or anti-unitary. This conclusion is also valid for higher dimensional linear representations (however, if the representation space is a Hilbert space, unitary or anti-unitary group elements will be quite different). The linear representations on Hermitian operator space are used to define the active operators.
 
For a general group $G$, if it has a nontrivial projective representation, which correspond to a SPT phase, then the active operators are defined in the following way: for a set of Hermitian operators $O^{\mathrm{ph}}_1,..., O^{\mathrm{ph}}_n$ acting on the physical spin Hilbert space, if we can find a set of Hermitian operators $O^{\mathrm{in}}_1,...,O^{\mathrm{in}}_n$ acting on the internal-spin Hilbert space (or the projective representation space), such that $O^{\mathrm{ph}}$ and $O^{\mathrm{in}}$ form the same $n$-dimensional real linear representation of $G$, then the operators $O^{\mathrm{ph}}$ are called active operators. Different SPT phases have different set of active operators, so we can use these active operators to distinguish different SPT phases.

\begin{table}[htbp]
\caption{Linear representations of $D_{2h}=D_{2}+T$}
\label{tab:Repd2h}
\begin{ruledtabular}
\begin{tabular}{c|cccccccc|ccc}
            & $E$        & $R_x$        & $R_y$      & $R_z$    & $T$ & $R_xT$       & $R_yT$     & $R_zT$&bases&operators&\\
\hline
$A_{g}$     &      1     &       1      &      1     &     1    &  1  &       1      &      1     &     1    &$|0,0\rangle$&$S_x^2,S_y^2,S_z^2$& \\
$B_{1g}$    &      1     &      -1      &     -1     &     1    &  1  &      -1      &     -1     &     1    &$i|1,z\rangle$&$S_{xy}$& \\
$B_{2g}$    &      1     &      -1      &      1     &    -1    &  1  &      -1      &      1     &    -1    &$i|1,y\rangle$&$S_{xz}$& \\
$B_{3g}$    &      1     &       1      &     -1     &    -1    &  1  &       1      &     -1     &    -1    &$i|1,x\rangle$&$S_{yz}$& \\
\hline
$A_{u}$     &      1     &       1      &      1     &     1    & -1  &      -1      &     -1     &    -1    &$i|0,0\rangle$&   $(S_{x,i}S_{yz,i+1})$&\\
$B_{1u}$    &      1     &      -1      &     -1     &     1    & -1  &       1      &      1     &    -1    &$|1,z\rangle$&$S_z$&\\
$B_{2u}$    &      1     &      -1      &      1     &    -1    & -1  &       1      &     -1     &     1    &$|1,y\rangle$&$S_y$& \\
$B_{3u}$    &      1     &       1      &     -1     &    -1    & -1  &      -1      &      1     &     1    &$|1,x\rangle$&$S_x$& 
\end{tabular}
\end{ruledtabular}
\end{table}

\begin{table}[htbp]
\caption{Linear representations of $\bar D_2=\{E,R_zT,R_xT,R_y\}$}
\label{tab:D2bar_linear}
\begin{ruledtabular}
\begin{tabular}{c|cccc|ccc}
         & $E$        & $R_zT$        & $R_xT$      & $R_y$&bases or operators&\\
\hline
$A$      &      1     &      1       &       1    &     1    &$|0,0\rangle$,$|1,y\rangle$&$S_y$,$S_x^2,S_y^2,S_z^2$\\
$B_1$    &      1     &      1       &      -1    &    -1    &$|1,x\rangle$,$i|1,z\rangle$&$S_x$,$S_{xy}$\\
$B_2$    &      1     &      -1      &      -1    &     1    &$i|0,0\rangle$,$i|1,y\rangle$& $S_{xz}$\\
$B_3$    &      1     &     -1       &       1    &    -1    &$|1,z\rangle$,$i|1,x\rangle$&$S_z$,$S_{yz}$
\end{tabular}
\end{ruledtabular}
\end{table}

\begin{table}[htbp]
\caption{Linear representations of $Z_2+T=\{E,R_z,T,R_zT\}$}
\label{tab:Z2+T_linear}
\begin{ruledtabular}
\begin{tabular}{c|cccc|cc}
         & $E$        & $R_z$        & $T$      & $R_zT$&bases or operators&\\
\hline
$A_g$      &      1     &      1       &       1    &     1    &$|0,0\rangle$,$i|1,z\rangle$& $S_{xy},S_x^2,S_y^2,S_z^2$\\
$A_u$    &      1     &      1       &      -1    &    -1    &$i|0,0\rangle$,$|1,z\rangle$&$S_z$\\
$B_g$    &      1     &     -1       &       1    &    -1    &$i|1,x\rangle$,$i|1,y\rangle$& $S_{yz},S_{xz}$\\
$B_u$    &      1     &     -1       &      -1    &     1    &$|1,x\rangle$,$|1,y\rangle$&$S_x,S_y$  
\end{tabular}
\end{ruledtabular}
\end{table}

\section{16 projective representations of $D_{2}+T$
group}\label{app:prjD2+T}

We have shown in appendices \ref{app:cohomology} and
\ref{app:revclassify} that the projective representations are
classified by the second group cohomology $H^2(G,U_T(1))$. However,
usually it is not easy to calculate the group cohomology. So we
choose to calculate the projective representations directly. In the
following we give the method through which we obtain all the 16
projective representations of $D_2+T$ in Table \ref{tab:prjRep2}.

The main trouble comes from the anti-unitarity of some symmetry
operators, such as the time reversal operator $T$. Under
anti-unitary operators (such as $T$), the matrix $A^m$ varies as
\begin{eqnarray}\label{MAMT}
\sum_{m'}u(T)_{mm'}(A^{m'})^*=M(T)^\dag A^m M(T).
\end{eqnarray}
Notice that $e^{i\alpha(T)}$ is absent because we can always set it
to be 1 by choosing proper phase of $A^m$. To see more difference
between the unitary operator and anti-unitary operators, we
introduce an unitary transformation to the bases of the virtual
`spin' such that $A^m$ becomes $\bar A^m=U^\dag A^mU$. Then for an
unitary symmetry operation $g$, Eq.~(\ref{MAM}) becomes
\begin{eqnarray*}
\sum_{m'}u(g)_{mm'}\bar A^{m'}=e^{i\alpha(g)}\bar M(g)^\dag \bar A^m
\bar M(g),
\end{eqnarray*}
where $\bar M(g)=U^\dag M(g) U$. However, for the anti-unitary
operator $T$,  $\bar A^m$ varies as
\begin{eqnarray*}
\sum_{m'}u(T)_{mm'}(\bar A^{m'})^*=\tilde M(T)^\dag \bar A^m \tilde
M(T),
\end{eqnarray*}
where $\tilde M(T)=U^\dag M(T)U^*=U^\dag [M(T)K]U$. Therefore, we
can see that $M(T)K$ as a whole is the anti-unitary projective
representation of $T$ when acting on the virtual `spin' space.

\begin{table}[htbp]
\caption{Unitary projective representations of $D_{2h}=D_2+T$, here
we consider $T$ as an unitary operator.} \label{tab:UnitprojD2h}
\begin{ruledtabular}
\begin{tabular}{c|cccc|}
&$R_z$&$R_x$&$T$\\
\hline
$A_{g}$     &     1      &      1     &      1      \\
$B_{1g}$    &     1      &     -1     &      1      \\
$B_{2g}$    &    -1      &     -1     &      1      \\
$B_{3g}$    &    -1      &      1     &      1      \\
$A_{u}$     &     1      &      1     &     -1      \\
$B_{1u}$    &     1      &     -1     &     -1      \\
$B_{2u}$    &    -1      &     -1     &     -1      \\
$B_{3u}$    &    -1      &      1     &     -1      \\
\hline
$E_1$       &     I      & $i\sigma_z$& $ \sigma_y$ \\
$E_2=E_1\otimes B_{3g}$       &    -I      & $i\sigma_z$& $ \sigma_y$  \\
\hline
$E_3$       & $ \sigma_z$&     I      & $i\sigma_y$ \\
$E_4=E_3\otimes B_{1g}$       & $ \sigma_z$&    -I      & $i\sigma_y$ \\
\hline
$E_5$       & $i\sigma_z$& $ \sigma_x$&     I       \\
$E_6=E_5\otimes A_{u}$       & $i\sigma_z$& $ \sigma_x$&    -I      \\
\hline
$E_7$       & $ \sigma_z$& $i\sigma_z$& $i\sigma_x$ \\
$E_8=E_7\otimes B_{1g}$       & $ \sigma_z$&-$i\sigma_z$& $i\sigma_x$ \\
\hline
$E_9$       & $i\sigma_z$& $ \sigma_x$& $i\sigma_x$ \\
$E_{10}=E_9\otimes A_{u}$    & $i\sigma_z$& $ \sigma_x$&-$i\sigma_x$\\
\hline
$E_{11}$    & $i\sigma_z$& $i\sigma_x$& $ \sigma_z$\\
$E_{12}=E_{11}\otimes B_{3g}$    &$i\sigma_z$& $i\sigma_x$& -$ \sigma_z$ \\
\hline
$E_{13}$    & $i\sigma_z$& $i\sigma_x$& $i\sigma_y$ \\
$E_{14}=E_{13}\otimes A_{u}$    & $i\sigma_z$& $i\sigma_x$&-$i\sigma_y$ \\
\end{tabular}
\end{ruledtabular}
\end{table}

The question is how to obtain the matrix $M(T)$. In
Ref.~\onlinecite{LLW1180}, we firstly treated $T$ as an unitary
operator, and we got 8 classes of unitary projective representations
for the group $D_{2h}$ (see Table~\ref {tab:UnitprojD2h}). By
replacing $M(T)$ by $M(T)K$, we obtained 8 different classes of
anti-unitary projective representations. However, not all the
projective representations can be obtained this way. Notice that
$[M(T)K]^2=1$ and $[M(T)K]^2=-1$ belong to two different projective
representations, the anti-unitary projective representations are
twice as many as the unitary projective representations.
Fortunately, all the remaining (anti-unitary) projective
representations can be obtained from the known ones. Notice that the
direct product of any two projective representations is still a
projective representation of the group, which can be reduced to a
direct sum of several projective representations. There may be new
ones in the reduced representations that are different from the 8
known classes. Repeating this procedure (until it closes), we
finally obtain 16 different classes of projective representations
(see appendix \ref{app:realizeSPT}). Notice that the Clebsch-Gordan
coefficients which reduce the product representation should be real,
otherwise it does not commute with $K$ and will not block
diagonalize the product representation matrix of $T$ (and other
anti-unitary symmetry operators). Because of this restriction, we
obtain four 4-dimensional irreducible projective representations
(IPRs) which are absent in the unitary projective representations.

\section{Realization of SPT phases in $S=1$ spin
ladders}\label{app:realizeSPT}

From the knowledge of section \ref{sec.chain}, together with
Eqs.~(\ref{MN}) and (\ref{Malpha}), we can construct different SPT
phases with spin ladders. From the discussion in section
\ref{sec.ladder.gen}, the projective representation $M(g)\otimes
N(g)$ is usually reducible. It can be reduced to several IPRs of the
same class. This class of projective representation determines which
phase the ladder belongs to. Thus, the decomposition of direct
products of different projective representations is important. Since
the SPT phases corresponding to $E_{13}, E_{11}, E_5, E_9$ ($T_0,
T_x, T_y, T_z$, separately) have been already realized in spin
chains, we will first study the decompositions of the direct product
of two of them.
\\
$E_5\otimes E_9=(\sigma_z, I, i\sigma_x)\oplus(\sigma_z, -I, i\sigma_x)=E_3'\oplus E_4'$;\\
$E_5\otimes E_{11}=(I,i\sigma_z,\sigma_x)\oplus(-I, i\sigma_z,\sigma_x)=E_1'\oplus E_2'$;\\
$E_5\otimes E_{13}=(\sigma_z,i\sigma_z,i\sigma_y)\oplus(\sigma_z,-i\sigma_z,i\sigma_y)=E_7'\oplus E_8'$;\\
$E_9\otimes E_{11}=(\sigma_z, i\sigma_z, i\sigma_x)\oplus(\sigma_z, -i\sigma_z,i\sigma_x)=E_7\oplus E_8$;\\
$E_9\otimes E_{13}=(I, i\sigma_z,\sigma_y)\oplus(-I,i\sigma_z,\sigma_y)=E_1\oplus E_2$;\\
$E_{11}\otimes E_{13}=(\sigma_z, I, i\sigma_y)\oplus(\sigma_z, -I,i\sigma_y)=E_3\oplus E_4$.\\

In above decomposition, all the CG coefficients are real. The three
matrices in each bracket are the representation matrices for the
three generators $R_z, R_x, T$, separately. We omitted the
anti-unitary operator $K$ for the representation matrix of $T$.
Further, $E_1$ and $E_2$ ($E_3$ and $E_4$, so on and so forth)
belong to the same class of projective representation, and differs
only by a phase transformation. So with spin ladders, we realize 6
SPT phases corresponding to the projective representations $E_1,
E_1', E_3, E_3', E_7, E_7'$.

Using these projective representations $E_1, E_1', E_3, E_3', E_7,
E_7'$, together with $E_{13}, E_{11}, E_5, E_9$,  we can repeat
above procedure and obtain more projective representations and their
corresponding SPT phases.
The result is shown below:
\\
 $E_1\otimes E_3=(\sigma_z, -i\sigma_z,i\sigma_x)\oplus(\sigma_z,i\sigma_z,-i\sigma_x)=E_7\oplus E_8$;\\
 $E_1\otimes E_5=(-I\otimes i\sigma_z, I\otimes i\sigma_x, -\sigma_y\otimes\sigma_z)=E_{11}'$;\\
 $E_1\otimes E_7=(\sigma_z, -I, i\sigma_y)\oplus(-\sigma_z,I,i\sigma_y)=E_3\oplus E_4$;\\
 $E_1\otimes E_9=(-i\sigma_z, -i\sigma_x, -i\sigma_y)\oplus(-i\sigma_z,i\sigma_x,-i\sigma_y)=E_{13}\oplus E_{14}$;\\
 $E_1\otimes E_{11}=(-I\otimes i\sigma_z, -I\otimes\sigma_x, \sigma_y\otimes I)=E_5'$;\\
 $E_1\otimes E_{13}=(-i\sigma_z, I, -i\sigma_x)\oplus(-i\sigma_z,-I,-i\sigma_x)=E_9\oplus E_{10}$;\\
\\
 $E_1'\otimes E_3=(-\sigma_z, -i\sigma_z,-i\sigma_y)\oplus(-\sigma_z,i\sigma_z,i\sigma_y)=E_7'\oplus E_8'$;\\
 $E_1'\otimes E_5=(-i\sigma_z, i\sigma_x,-\sigma_z)\oplus(-i\sigma_z,i\sigma_x,\sigma_z)=E_{11}\oplus E_{12}$;\\
 $E_1'\otimes E_7=(-\sigma_z, I, i\sigma_x)\oplus(-\sigma_z,-I,-i\sigma_x)=E_3'\oplus E_4'$;\\
 $E_1'\otimes E_9=(-I\otimes i\sigma_z, I\otimes i\sigma_x, -i\sigma_y\otimes\sigma_y)=E_{13}'$; \\
 $E_1'\otimes E_{11}=(-i\sigma_z, -\sigma_x,I)\oplus(-i\sigma_z,-\sigma_x,-I)=E_5\oplus E_6$;\\
 $E_1'\otimes E_{13}=(-I\otimes i\sigma_z, -I\otimes \sigma_x, -i\sigma_y\otimes \sigma_x)=E_9'$;\\
\\
 $E_3\otimes E_5=(-I\otimes i\sigma_z, I\otimes \sigma_x, i\sigma_y\otimes\sigma_x )=E_{9}'$;\\
 $E_3\otimes E_7=( -I, i\sigma_x, \sigma_y)\oplus(I,i\sigma_x,\sigma_y)=E_1\oplus E_2$;\\
 $E_3\otimes E_9=(-I\otimes i\sigma_z, I\otimes \sigma_x, -\sigma_y\otimes I)=E_{5}'$;\\
 $E_3\otimes E_{11}=(-i\sigma_z,i\sigma_x,i\sigma_y)\oplus(-i\sigma_z,i\sigma_x,-i\sigma_y)=E_{13}\oplus E_{14}$;\\
 $E_3\otimes E_{13}=(-i\sigma_z, i\sigma_x, -\sigma_z)\oplus(-i\sigma_z,-i\sigma_x,\sigma_z)=E_{11}\oplus E_{12}$;\\
\\
 $E_3'\otimes E_5=(-i\sigma_z,\sigma_x,-i\sigma_x)\oplus(-i\sigma_z,-\sigma_x,-i\sigma_x)=E_{9}\oplus E_{10}$;\\
 $E_3'\otimes E_7=( -I, i\sigma_x, -\sigma_z)\oplus(I,i\sigma_x,-\sigma_z)=E_1'\oplus E_2'$;\\
 $E_3'\otimes E_9=(-i\sigma_z, \sigma_x, -I)\oplus(-i\sigma_z,-\sigma_x,I)=E_{5}\oplus E_{6}$;\\
 $E_3'\otimes E_{11}=(-I\otimes i\sigma_z, I\otimes i\sigma_x, i\sigma_y\otimes\sigma_y)=E_{13}'$;\\
 $E_3'\otimes E_{13}=(-I\otimes i\sigma_z, I\otimes i\sigma_x,  \sigma_y\otimes\sigma_z)=E_{11}'$;\\
\\
 $E_7\otimes E_{5}=(-I\otimes i\sigma_z, I \otimes i\sigma_x , -\sigma_y\otimes\sigma_y)=E_{13}'$;\\
 $E_7\otimes E_9=(-i\sigma_z, i\sigma_x, \sigma_z)\oplus(i\sigma_z,-i\sigma_x,\sigma_z)=E_{11}\oplus E_{12}$;\\
 $E_7\otimes E_{11}=(-i\sigma_z,  \sigma_x, -i\sigma_x)\oplus(i\sigma_z,\sigma_x,i\sigma_x)=E_{9}\oplus E_{10}$;\\
 $E_7\otimes E_{13}=(-I\otimes i\sigma_z, -I \otimes \sigma_x , -\sigma_y\otimes I)=E_{5}'$;\\
\\
 $E_7'\otimes E_{5}=(-i\sigma_z,i\sigma_x,i\sigma_y)\oplus(-i\sigma_z,i\sigma_x,-i\sigma_y)=E_{13}\oplus E_{14}$;\\
 $E_7'\otimes E_9=(-I\otimes \sigma_z, I\otimes i\sigma_x, \sigma_y\otimes\sigma_z)=E_{11}'$;\\
 $E_7'\otimes E_{11}=(-I\otimes i\sigma_z, -I\otimes \sigma_x, -i\sigma_y\otimes\sigma_x)=E_{9}'$;\\
 $E_7'\otimes E_{13}=(-i\sigma_z, \sigma_x, -I)\oplus(-i\sigma_z,\sigma_x,I)=E_5\oplus E_6$;\\
\\
 $E_1\otimes E_1'= (I,I,\sigma_y)\oplus(I,-I,-\sigma_y)=E_0'\oplus E_0'$;\\
 $E_3\otimes E_3'= (-I,I,-\sigma_y)\oplus(I,I,\sigma_y)=E_0'\oplus E_0'$;\\
 $E_7\otimes E_7'= (I,-I,\sigma_y)\oplus(-I,I,\sigma_y)=E_0'\oplus E_0'$.\\

Above we get four SPT phases corresponding to $E_5', E_9', E_{11}',
E_{13}'$, all of them have 4-dimensional end `spins'. We also get a
SPT phase corresponding to $E_0'$, which has 2-dimensional end
`spins'.

Notice that the number of classes of unitary projective
representations of $D_{2h}$ is 8, but considering that $T$ is
anti-unitary such that $T^2$ can be either 1 or -1, we obtain 16
classes of projective representations for $D_2+T$.


\begin{thebibliography}{10}
\bibitem{note:SO(3)} When $a=b=c=1$, this state is invariant under
$SO(3)+T$, where $SO(3)$ is generated by $S_x, S_y, S_z$ and
$T=e^{i\pi S_y}K$. From Ref.~\onlinecite{CGW1123}, systems with
$SO(3)+T$ symmetry have 4 SPT phases. It seems strange that its
subgroup $D_2+T$ contains more SPT phases. Actually, there are four
distinct $SO(3)+T$ groups which contian $D_2+T$ as a subgroup. In
these four groups, $T$ is always defined as $T=e^{i\pi S_y}K$, but
the $SO(3)$ parts are different. Except for the one mentioned above,
we have additional three choices: $-S_x, S_{xz}, S_{xy}$ or $S_{yz},
-S_y, S_{xy}$ or $S_{yz}, S_{xz}, -S_z$. Each of the four groups
contains 4 SPT phases, so their common subgroup $D_2+T$ contains
$4\times4=16$ SPT phases.
\bibitem{laddernote}Actually, provided that the the symmetry group of the Hamiltonian of the ladder is
$D_2+T$, inter-chain interactions must be considered (otherwise the
symmetry group should be $(D_2+T)\otimes(D_2+T)$). Here we take the
limit that the strength of inter-chain interaction tends to zero.
\end{thebibliography}

%

\end{document}